\documentclass[article,twocolumn,pra,showpacs,superscriptaddress]{revtex4-1}
\usepackage[latin1]{inputenc}
\usepackage{psfrag}
\usepackage{bm}
\usepackage{multirow,amssymb,amsmath,amsfonts}
\usepackage{graphicx}
\usepackage{color}

\newcommand{\comments}[1]{}

\newcommand{\bfsig}{\boldsymbol{\hat{\sigma}}}
\newcommand{\bflam}{\boldsymbol{\lambda}}
\newcommand{\hr}{g_\textrm{r}}
\newcommand{\hi}{g_\textrm{i}}
\newcommand{\h}{g}

\newcommand{\fp}{f_{+}}
\newcommand{\fm}{f_{-}}
\newcommand{\fpm}{f_{\pm}}

\newcommand{\N}{\mathcal{N}}
\newcommand{\Nana}{\mathcal{N}^{\text{ana}}}
\newcommand{\Gammar}{\Gamma_{\textrm{r}}}
\newcommand{\Gammai}{\Gamma_{\textrm{i}}}
\newcommand{\trel}{\tau_{r}}
\newcommand{\tcor}{\tau_{c}}
\newcommand{\tsys}{\tau_{s}}

\newcommand{\omegac}{\omega_c}
\newcommand{\omegaa}{\omega_A}
\newcommand{\x}{\mu}
\newcommand{\PhiSA}{\Phi_{\text{SA}}}
\newcommand{\PhiRWA}{\Phi_{\text{RWA}}}
\newcommand{\JL}{J_{\text{L}}}
\newcommand{\JO}{J_{\text{O}}}

\newcommand{\Tr}{\textrm{Tr}}

\newcommand{\rhoE}{\hat{\rho}_E}

\newcommand{\rhoSI}{\hat{\rho}}
\newcommand{\HS}{\hat{H}_S}
\newcommand{\HE}{\hat{H}_E}

\newcommand{\Hone}{\hat{H}_I}
\newcommand{\HoneI}{\hat{H}^{\text{Int}}}

\newcommand{\sig}{\hat{\sigma}}
\newcommand{\sigx}{\hat{\sigma}_x}
\newcommand{\sigy}{\hat{\sigma}_y}
\newcommand{\sigz}{\hat{\sigma}_z}

\newcommand{\sigp}{\hat{\sigma}_{+}}
\newcommand{\sigm}{\hat{\sigma}_{-}}

\begin{document}

\title{Effects of the  rotating-wave and secular approximations on non-Markovianity}
\author{H. M\"akel\"a}
\email{harri.makela@aalto.fi}
\affiliation{QCD Labs, COMP Centre of Excellence, Department of Applied Physics, Aalto University,
P.O. Box 13500, FI-00076 AALTO, Finland}
\author{M. M\"ott\"onen}
\affiliation{QCD Labs, COMP Centre of Excellence, Department of Applied Physics, Aalto University,
P.O. Box 13500, FI-00076 AALTO, Finland}
\affiliation{Low Temperature Laboratory (OVLL), Aalto University, P.O. Box 13500, FI-00076 Aalto, Finland}

\begin{abstract}
We study the effect of the rotating-wave approximation (RWA) and the secular approximation (SA)
on the non-Markovian behavior in the spin-boson model at zero-temperature. We find that both the RWA and SA lead to a dramatic reduction in the observed non-Markovianity. In general, non-Markovian dynamics is observed for the whole relaxation time of the system, whereas the RWA and SA lead to such dynamics only on the short time scale of the environmental correlation time. Thus, the RWA and SA are not necessarily justified in the studies of non-Markovianity although they can estimate the state of the system precisely. Furthermore, we derive an accurate analytical expression for the non-Markovianity measure without the RWA or SA. This expression yields important insight into the physics of the problem.
\end{abstract}


\maketitle
\section{Introduction}
 Coupling between an open quantum system and its environment is typically assumed to lead to Markovian processes in which information
 flowing from the open system to the environment never returns~\cite{Breuer,Weiss}. Although this is the most commonly studied scenario, non-Markovian processes can have a prominent role under certain conditions.
 This can be the case, for example, if the system--environment coupling is strong, the reservoir is structured or has a finite size, or
 the temperature is low. Detailed characterization of non-Markovian processes is therefore called for, but
difficulties in finding a proper definition for quantum non-Markovianity have complicated this task considerably. Due to recent advancements in the definition and quantification of quantum non-Markovianity~\cite{Wolf08,Breuer09,Rivas10},  interest in non-Markovian dynamics of open quantum systems has increased rapidly. For an overview of topics discussed in this context see, for example,
Refs.~\cite{Laine10,Chruscinski10,Lu10,Chruscinski11,Apollaro11,Haikka11,Znidaric11,Liu11,Mazzola12,
Rodriguez12,Zeng12,Clos12,Zhang12,Wissmann12,Bylicka13}.

A commonly used approximation in the theory of open quantum systems is the removal of terms that oscillate
fast with respect to some characteristic time scales of the system.
There are two distinct ways to implement this type of an approximation.
One is based on dropping the rapidly oscillating terms from the interaction picture Hamiltonian~\cite{Walls,Scully},
whereas in the other approach these terms are removed from the interaction picture master equation for the reduced density operator~\cite{Breuer,Agarwal}.
In the following we refer to the former as the rotating-wave approximation (RWA) and to the latter as the secular approximation (SA)~\cite{Breuer}.
In the theory of open quantum systems, the SA is often used in combination with the Markov approximation  to render the master equation in the Lindblad form.
The validity of the RWA has been studied by many authors, while that of the SA seems to have received less attention.

In the early 1970s Agarwal studied spontaneous emission from a collection of identical
two-level systems, finding that the RWA gives an erroneous frequency shift in the energy levels of the two-level
systems~\cite{Agarwal,Agarwal71,Agarwal73}. This problem does not arise if the SA is used~\cite{Agarwal73}. Ford and O'Connell discovered that, in general, the energy spectrum of an RWA Hamiltonian is not bounded from below~\cite{Ford97}.
Intravaia {\it et al.} found that the terms neglected in the RWA and SA may have experimentally measurable effects on the dynamics of a quantum harmonic oscillator~\cite{Intravaia03}.
In 2008, Zheng {\it et al.} showed that the Zeno time of the quantum Zeno effect is two orders of magnitude longer and the anti-Zeno effect
 disappears if the RWA is not used~\cite{Zheng08}. In 2012 Larson argued that the Berry phase appearing in certain cavity quantum electrodynamics settings
can be considered an artifact resulting from the use of the RWA~\cite{Larson12}.
In the same year, Peano {\it et al.}  found that the tunneling rate of a
parametrically modulated oscillator can be exponentially increased by processes caused by terms disregarded in the RWA~\cite{Peano12}.

In Refs.~\cite{Pekola10,Solinas10}, it was discovered that the SA yields unphysical results in the context of Cooper pair pumping. The SA has also been shown to lead to the breaking of very general conservation laws  of some observables~\cite{Salmilehto12} and to spurious effects in the context of environment-assisted entanglement generation~\cite{Benatti10}. Further discussion on the RWA and SA and more references are provided in Ref.~\cite{Fleming10}. However, no detailed study on the effect of RWA and SA on non-Markovian behavior exists. Is non-Markovianity in general as vulnerable to SA as the conservation of electric charge in the Cooper pair pump~\cite{Pekola10,Solinas10,Salmilehto12}?

In this paper, we consider the effects of the RWA and SA on non-Markovian processes occurring
in the spin boson model at zero temperature.
We quantify non-Markovianity using the trace-distance-based measure defined in Ref.~\cite{Breuer09} and
employ a master equation which has been derived without using the RWA or SA.
The non-Markovianity arising from this type of a master equation has been discussed previously in Refs.~\cite{Clos12,Zeng12}.
In Ref.~\cite{Clos12}, the behavior of the trace-distance-based non-Markovianity measure as a function of the temperature of the environment and the parameters of the spectral density was studied in the context of the spin-boson model. However, the effects of the RWA and SA were not considered, which prevented the authors from arriving at the conclusions of this work.
In \cite{Zeng12}, the trace-distance-based non-Markovianity measure was compared with the
divisibility-based measure (see~\cite{Rivas10}) in the spin-boson model at zero temperature.
However, the optimization over initial-state pairs appearing in the definition of the
non-Markovianity measure was not carried out. In this paper,
we perform the optimization and find out that the value of the measure can be orders of magnitude greater than previously predicted. Strikingly, we show that the RWA and SA lead to a dramatic reduction in the value of the measure. This is because the rapidly oscillating terms omitted in these approximations are responsible for the major part of non-Markovian dynamics. Moreover, we illustrate that if these rapidly oscillating terms are not dropped, an analytical expression for the value of the non-Markovianity measure can be derived straightforwardly. With the help of this expression, the properties of the non-Markovianity measure can be studied without having to explicitly solve the dynamics of the system.

This paper is organized as follows. In Sec.~\ref{sec_Hamiltonian}, we define the Hamiltonian,
present the second-order master equations, and define the spectral densities used in this work.
The dynamical map giving the time evolution of the two-level system and the complete positivity of this map are discussed in Sec.~\ref{sec_DM}.
In Sec.~\ref{sec_RWA_and_SA}, we present the non-Markovianity measure and
calculate its value using the full master equation and the master equations obtained using the RWA and SA. We find that in the latter two cases the value of the measure can be orders of magnitude smaller than in the former case, which is the main result of our work. We explain the reason for this reduction in non-Markovianity in Sec.~\ref{sec_Full_ME}, where we derive an analytical formula for the non-Markovianity measure. We show numerically that this formula estimates the value of the non-Markovianity measure very precisely. With the help of the analytical expression, we  show that for certain values of the parameters of the spectral density the dynamics is Markovian. We also discuss the validity of the SA and the effects of the Markov approximation.
Finally, in Sec.~\ref{sec_conclusions}, we summarize our results.

\section{Hamiltonian and master equations}
\label{sec_Hamiltonian}
We consider a two-level atom with  energy level separation $\omegaa$ coupled to
an environment consisting of harmonic oscillators. We denote the eigenbasis of the two-level
system by $\{|0\rangle,|1\rangle\}$ and define $\hat{\sigma}_x=|0\rangle\langle 1|+
|1\rangle\langle 0|,\hat{\sigma}_y=i|0\rangle\langle 1|-i
|1\rangle\langle 0|$, and $\hat{\sigma}_z=|1\rangle\langle 1|-|0\rangle\langle 0|$.
The total Schr\"odinger picture Hamiltonian reads
\begin{align}
\hat{H}^{\text{Sch}} = \frac{\omegaa}{2}\sigz +  \sum_k\omega_k \hat{b}_k^\dagger \hat{b}_k +
\sum_k \sigx (g_k\hat{b}_k + g_k^*\hat{b}^\dagger_k),
\end{align}
where $*$ indicates the complex conjugate and the terms on the right-hand side give the system, environment, and interaction Hamiltonians $\HS$, $\HE$, and $\Hone$, respectively. The index $k$ labels the modes of the environment, $\omega_k$ is the frequency
 of the $k$th oscillator, $g_k$ is a mode-dependent coupling constant, and $[\hat{b}_k,\hat{b}_{l}^{\dag}] = \delta_{kl}$. We set $\hbar=1$ throughout the paper.
 
 In the interaction picture  with respect to $\HS+\HE$, the  Hamiltonian becomes
\begin{align}
\label{eq_H_IP}
\HoneI(t) =&\sum_k g_k [e^{i(\omegaa-\omega_k)t}\sigp + e^{-i(\omegaa+\omega_k)t}\sigm]\hat{b}_k  +\textrm{H.c.} ,
\end{align}
where H.c. denotes the Hermitian conjugate
and $\hat{\sigma}_{\pm}=(\hat{\sigma}_x\pm i\hat{\sigma}_y)/2$.
In the limit of a weak interaction between
the open system and the environment, the dynamics is well described by the second-order
 time-convolutionless master equation  \cite{Breuer}.
We assume that the initial state of the total system factorizes as $\rhoSI(0)\otimes\rhoE^{\text{vac}}$, where $\rhoSI(0)$ is the initial state of the two-level system and
$\rhoE^{\text{vac}}$ is the vacuum state of the environment.
The interaction picture master equation reads
\begin{align}
\label{eq_ME}
\nonumber
&\frac{d}{dt}\rhoSI(t) = i\frac{1}{2}\hi (t) [\sigz,\rhoSI(t)]\\
\nonumber
& +\sum_{k=\pm}f_k (t) [\sig_k\rhoSI(t)\sig_k^\dag
-\frac{1}{2}\{\sig_k^\dag\sig_k,\rhoSI(t)\}]\\
&+ e^{i2\omegaa t}\h (t) \sigp\rhoSI(t)\sigp +  e^{-i2\omegaa t}\h^ * (t)\sigm\rhoSI(t)\sigm,
\end{align}
where $\hi$ is the imaginary part of $\h$ and
\begin{align}
\label{eq_fpm}
f_\pm(t) &=2\int_{0}^{\infty}d\omega \int_{0}^t ds\  J(\omega)\cos[(\omega\pm\omegaa)s],\\
\label{eq_h}
\h(t) &=2\int_{0}^{\infty}d\omega \int_{0}^t ds\  J(\omega)\cos(\omega s)
e^{-i\omegaa s}.
\end{align}
Here $J$ is the spectral density of the environmental modes.
Note that for the real part of $\h$, denoted by $\hr$, the equation $\hr=(f_++f_-)/2$ holds.

The RWA corresponds to dropping the terms multiplied
by $e^{\pm i(\omegaa+\omega_k)t}$ in Eq. \eqref{eq_H_IP} and yields the RWA master equation,
\begin{align}
\label{eq_ME_RWA}
\nonumber
&\frac{d}{dt}\rhoSI(t) = i\frac{1}{2}h(t)[\sigz,\rhoSI(t)]\\
& +f_{-} (t) [\sigm\rhoSI(t)\sigp -\frac{1}{2}\{\sigp\sigm,\rhoSI(t)\}],
\end{align}
where $h(t)=\int_{0}^{\infty}d\omega \int_{0}^t ds\  J(\omega)\sin[(\omega-\omegaa) s]$.
Although the dynamics determined by the RWA Hamiltonian can be solved exactly~\cite{Breuer}, we have employed the second-order approach here in order to consistently compare the resulting dynamics with those given by Eq.~\eqref{eq_ME}. The difference between the exact and second-order RWA dynamics is very small in the weak-coupling limit considered here. 

The SA master equation can be obtained by removing the rapidly oscillating terms from Eq. \eqref{eq_ME} as described in Ref.~\cite{Breuer}. The calculation of the SA master equation is straightforward and yields
\begin{align}
\label{eq_ME_SA}
\nonumber
&\frac{d}{dt}\rhoSI(t) =
i\frac{1}{2}\hi (t) [\sigz,\rhoSI(t)]\\
&+\sum_{k=\pm}f_k (t) [\sig_k\rhoSI(t)\sig_k^\dag-\frac{1}{2}\{\sig_k^\dag\sig_k,\rhoSI(t)\}].
\end{align}
It is possible to eliminate the first term on the right-hand side of Eqs.~\eqref{eq_ME},~\eqref{eq_ME_RWA}, and~\eqref{eq_ME_SA} by a change of basis~\cite{footnote}.

Three important time scales characterizing the dynamics are the environment correlation time $\tcor$,
the relaxation time $\trel$, and the characteristic time scale of the
intrinsic evolution of the two-level system
\begin{align}
\tsys=\frac{1}{\omegaa}.
\end{align}
The relaxation time scale is inversely proportional to the strength of the system-environment interaction. The precise definition is given in Eq.~\eqref{eq_trel} below.
The environment correlation time $\tcor$ describes the time after which $\fpm$ and $\h$ have reached their asymptotic values: $\fpm(t\gg \tcor)\approx\fpm(\infty)\geq 0$ and $\h(t\gg\tcor)\approx \h(\infty)$. If $t\gg\tcor$, the RWA and SA master equations are in the Lindblad form, and the dynamics is Markovian. However, as we show below, non-Markovian dynamics is possible at times  $t\gg\tcor$ if the full master equation~\eqref{eq_ME} is used.

In this paper, we consider Lorentzian and Ohmic spectral densities.
The former is defined as
\begin{align}
\JL(\omega)= \frac{\alpha}{2\pi} \frac{\lambda^2}{(\omega+\Delta-\omegaa)^2 +\lambda^2},
\end{align}
where $\Delta$ is the detuning from the energy separation of the two-level system and
$\lambda>0$ characterizes the spectral width of the coupling and determines
the environment correlation time as $\tcor=1/\lambda$.
The Ohmic spectral density with Lorentz-Drude cutoff reads
\begin{align}
\label{eq_JO}
\JO(\omega) = \frac{\alpha}{\pi}\frac{\omega}{\omegaa} \frac{\omegac^2}{\omega^2+\omegac^2},
\end{align}
where $\omegac$ is the cutoff frequency. The environment correlation time is $\tcor =1/\omegac$.
In both spectral densities $\alpha$ characterizes the strength of the system-environment interaction.
The frequency integration is restricted to non-negative frequencies in Eqs. \eqref{eq_fpm} and
\eqref{eq_h}. However, in the case of $\JL$,  the lower limit can be extended to $-\infty$ as we choose the values of $\omegaa,\lambda$, and $\Delta$ in such a way that the contribution to the integrals from negative frequencies is negligible compared with the contribution arising from positives frequencies.
We show in Appendix~\ref{appendix_Functions} that this is the case if $\omegaa-\Delta\gg \lambda$. 
Expressions for $\fpm$ and $\h$ corresponding to the Lorentzian  and Ohmic
spectral density are given in Appendix~\ref{appendix_Functions}.

\section{Dynamical Map}
\label{sec_DM}
Before proceeding to quantify non-Markovianity, we define a map that gives the time evolution of the system. This will be used later in the calculation of the non-Markovianity measure.
The solution of Eq.~\eqref{eq_ME}  can be given in terms of a one-parameter family of maps $\Phi=\{\Phi_t\ |\ t\geq 0\}$, so that $\hat{\rho}(t)=\Phi_t[\hat{\rho}(0)]$.
We write the state of the two-level system as
\begin{align}
\hat{\rho}(t)=
\begin{pmatrix}
\rho_{11}(t) & \rho_{10}(t)\\
\rho_{01}(t) & \rho_{00}(t)
\end{pmatrix},
\end{align}
where $\rho_{ij}=\langle i|\hat{\rho}|j\rangle$.
The solution of Eq.~\eqref{eq_ME} can be expressed as
\begin{align}
\label{eq_rho11t}
\rho_{11}(t) &=\frac{1}{2}\left\{1+u(t)+ e^{-\Gammar(t)}\left[2\rho_{11}(0)-1\right]\right\}, \\
\label{eq_rho00t}
\rho_{00}(t) &=\frac{1}{2}\left\{1-u(t)+ e^{-\Gammar(t)}\left[2\rho_{00}(0)-1\right]\right\}, \\
\label{eq_rho10t}
\rho_{10}(t) &= e^{-\frac{1}{2}\Gamma^*(t)}[ v_{1}(t) \rho_{10}(0)+v_{2}(t)\rho_{01}(0)],\\
\label{eq_rho01t}
\rho_{01}(t) &= e^{-\frac{1}{2}\Gamma(t)}[ v_{1}^*(t) \rho_{01}(0)+v_{2}^*(t)\rho_{10}(0)],
\end{align}
where
\begin{align}
\label{eq_Gamma}
\Gamma(t) &=\Gammar(t)+i\Gammai(t)=2\int_{0}^{t} ds\ \h(s)
\end{align}
and
$u(t) =\int_{0}^{t} ds\ e^{\Gammar(s)-\Gammar(t)} [\fp(s)-\fm(s)].$
Functions $v_1$ and $v_2$  are obtained as the solution of the differential equation
\begin{align}
\label{eq_v}
\frac{d}{dt}v_{1,2}(t) &= \h(t)e^{i[2\omegaa t-\Gammai(t)]}  v^*_{2,1}(t),
\end{align}
with the initial conditions $v_{1}(0)=1,v_{2}(0)=0$.
In this paper, we solve Eq.~\eqref{eq_v} numerically.
From Eqs.~\eqref{eq_rho11t}-\eqref{eq_rho01t} we observe that the
decay rate of the elements of the density matrix is determined by $\Gammar$.
Since $\Gammar(t\gg\tcor)\approx 2\hr(\infty) t$,
we define the relaxation time as
\begin{align}
\label{eq_trel}
\trel =\frac{1}{\hr(\infty)}.
\end{align}
The map $\Phi_t$ giving the state at time $t$ can be given as
\begin{align}
\label{eq_dynmap}
\hat{\rho}(t)=\sum_{i=1}^4 \Lambda_i(t) \hat{A}^\dagger_i(t)\hat{\rho}(0)\hat{A}_i(t),
\end{align}
where
\begin{align}
\label{eq_Lambda12}
\Lambda_{1,2}(t) &=\frac{1+e^{-\Gammar(t)} \mp \sqrt{4|v_1(t)|^2e^{-\Gammar(t)}+u(t)^2}}{4},\\
\label{eq_Lambda34}
\Lambda_{3,4}(t) &=\frac{1-e^{-\Gammar(t)}\mp \sqrt{4|v_2(t)|^2e^{-\Gammar(t)}+u(t)^2}}{4},
\end{align}
and the labeling is such that an odd (even) subscript corresponds to $-$ $(+)$. The operators $\{\hat{A}_i\}$ are defined in appendix~\ref{appendix_Dynamical_map}. 
To characterize a physically well-defined time evolution, the map $\Phi_t$ should
be completely positive and trace preserving (CPT) for any $t\geq 0$.
Note that complete positivity implies that $\hat{\rho}(t)$ is positive. 
The map $\hat{\Phi}_t$ is CPT if and only if $\Lambda_i(t)\geq 0$, $i=1,2,3,4$, and $\sum_{i=1}^{4} \Lambda_i(t)\hat{A}_i(t)\hat{A}^\dagger_i(t)=\mathbb{I}_2$, where $\mathbb{I}_2=|0\rangle\langle 0|+|1\rangle\langle 1|$. Our numerical checks indicate that $\Phi_t$ is CPT for any $t$ and any values of the parameters used in this paper.

\section{Non-Markovian dynamics and the rotating-wave and secular approximations}
\label{sec_RWA_and_SA}
In this section, we define the non-Markovianity measure used in this paper and calculate its value using the RWA and SA master equations. We compare these values to the ones obtained using the full master equation given in Eq.~\eqref{eq_ME}. The non-Markovian dynamics obtained using Eq.~\eqref{eq_ME} will be discussed in more detail in Sec.~\ref{sec_Full_ME}.
\subsection{Definition of the non-Markovianity measure}
We quantify non-Markovianity using the measure presented in Ref. \cite{Breuer09}.
According to the definition of this measure, Markovian processes lead to the reduction of the distinguishability of physical states, whereas non-Markovian processes increase the distingushability. The distinguishability of arbitrary states $\rhoSI_1$ and $\rhoSI_2$ at time $t$ can be quantified by the trace distance
\begin{align}
D[\rhoSI_1(t),\rhoSI_2(t)] =\frac{1}{2}\|\rhoSI_1(t)-\rhoSI_2(t)\|_1,
\end{align}
where $\|\hat{A}\|_1=\Tr\sqrt{\hat{A}^\dagger\hat{A}}$.
An open quantum system is said to exhibit non-Markovian behavior if for some pair of initial states
$\left(\rhoSI_1(0),\rhoSI_2(0)\right)$ the distance between $\rhoSI_1(t)$ and
$\rhoSI_2(t)$ increases at some $t$, indicating an increase in the distinguishability:
\begin{align}
\sigma(\rhoSI_{1,2};t) =\frac{d}{dt} D\left[\rhoSI_1(t),\rhoSI_2(t)\right] > 0.
\end{align}
Based on this definition of non-Markovian dynamics, the total amount of non-Markovianity can be quantified by
\begin{align}
\label{eq_N}
\N(\Phi) = \max_{\substack{\rhoSI_{1,2}(0)}}\frac{1}{2}\int_{0}^{\infty} dt
\Big[|\sigma(\rhoSI_{1,2};t)|+\sigma(\rhoSI_{1,2};t)\Big],
\end{align}
which includes a maximization over all possible initial-state pairs $(\rhoSI_{1}(0),\rhoSI_{2}(0))$.
The calculation of $\N(\Phi)$ can be simplified by noting that it is sufficient to choose the two initial states to be orthogonal states that lie on the boundary of the space of physical states \cite{Wissmann12}.

\subsection{RWA and SA master equations}
We write the state of the two-level system as
$\hat{\rho}=(\mathbb{I}_2+\bflam\cdot\bfsig)/2$, where the length of the Bloch vector $\bflam$ is restricted by the equation $0\leq\|\bflam\|=\sqrt{\bflam\cdot\bflam}\leq 1$.
Two necessary conditions for the pair $(\hat{\rho}_1(0),\hat{\rho}_2(0))$
to maximize the non-Markovianity measure are that the Bloch vectors of these states, denoted by $\bflam^1(0)$ and $\bflam^2(0)$, fulfill the conditions $\bflam^2(0)=-\bflam^1(0)$ and
$\|\bflam^1(0)\|=\|\bflam^2(0)\|=1$ (see Ref.~\cite{Wissmann12}).
As we show in Appendix \ref{appendix_N}, for the RWA and SA master equations
the  non-Markovianity is maximized by choosing $\bflam^1(0)=-\bflam^2(0)=(0,0,1)$.
A similar result has been previously obtained numerically in the RWA case in Ref.~\cite{Breuer09}.
The value of the non-Markovianity measure reads here
\begin{align}
\label{eq_NRWA}
\N(\PhiRWA)=\frac{1}{2}\int_{0}^{\infty} dt\ [|\fm(t)|-\fm(t)]  e^{-\Gamma^{\textrm{RWA}}(t)},
\end{align}
where $\Gamma^{\textrm{RWA}}(t)=\int_{0}^{t} ds\fm(s)$.
For the SA master equation we find that (see Appendix \ref{appendix_N})
\begin{align}
\label{eq_NSA}
\N(\PhiSA)&=\int_{0}^{\infty} dt\ [|\hr(t)|-\hr(t)]  e^{-\Gammar(t)},
\end{align}
where, as mentioned after Eq.~\eqref{eq_h}, $\hr=(f_+ + f_-)/2$.

Here and in what follows $\N_\text{L}$  and $\N_\text{O}$ refer to the non-Markovianity measure
calculated using the Lorentzian and Ohmic spectral densities, respectively.
For the Lorentzian spectral density, $\fp(t)$ approaches zero with increasing $\omegaa$, so that  $\lim_{\tsys\to 0}\N_\text{L}(\PhiSA)=\N_\text{L}(\PhiRWA)$.

\begin{figure}[t]
\begin{center}
\includegraphics[scale=1.0]{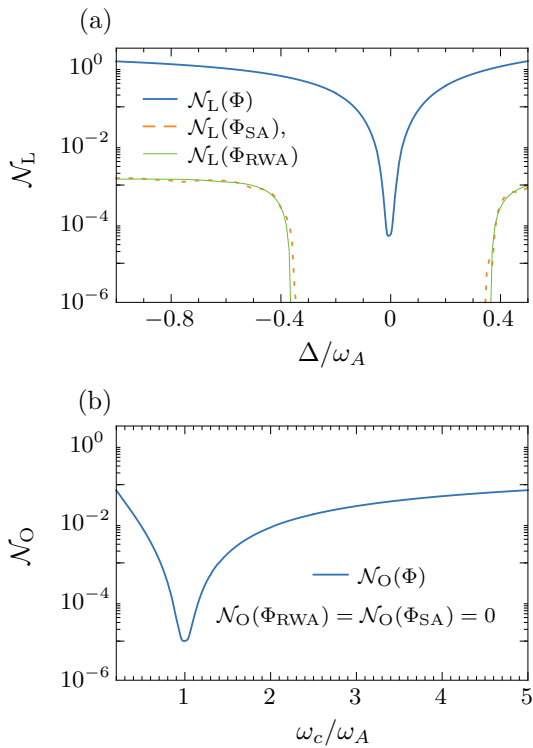}
\end{center}
\caption{
(Color online) Numerically calculated non-Markovianity corresponding to the (a) Lorentzian and (b) Ohmic  spectral densities. In (a), the top solid line is obtained using the full master equation [see Eq.~\eqref{eq_ME}], while the bottom solid and dashed lines show the non-Markovianity obtained under the RWA and SA, respectively  [see Eqs.~\eqref{eq_NRWA} and~\eqref{eq_NSA}].
In (b), the dynamics is Markovian if the RWA or SA master equation is used.
The reason for the large difference between $\mathcal{N}(\Phi)$, and $\mathcal{N}(\Phi_\text{SA})$ and $\mathcal{N}(\Phi_\text{RWA})$ is explained in Sec.~\ref{sec_Full_ME}.
Here we have chosen $\alpha/\omegaa=0.01$, and in (a) $\lambda/\omegaa=\tsys/\tcor=0.1$.
\label{fig_comparison}}
\end{figure}

The behavior of $\N_\text{L}(\PhiSA)$ and $\N_\text{L}(\PhiRWA)$ as a function of the detuning of the Lorentzian spectral density is plotted in Fig.~\ref{fig_comparison}(a) together with the value of the non-Markovianity measure obtained using the full master equation [see Eq.~\eqref{eq_ME}]. Strikingly, the full approach yields orders of magnitude greater non-Markovianty compared with that obtained using the RWA or SA.
For the Ohmic spectral density, the destructive effect of the RWA and SA is even more pronounced: the dynamics is Markovian if the RWA or SA is used,
but the full master equation yields non-Markovian dynamics, as shown in Fig.~\ref{fig_comparison}(b).
In the case of $\N_\text{L}(\Phi)$ and $\N_\text{O}(\Phi)$, we determine the initial-state pair maximizing the value of the non-Markovianity measure numerically.

The huge differences in the values of the non-Markovianity measures caused by RWA and SA can be explained by comparing the lengths of the time intervals during which
non-Markovian dynamics takes place. In the case of the RWA and SA master equations, the length of
this time interval is proportional to the correlation time $\tcor$, whereas in the case of the full master equation, it is proportional to the relaxation time $\trel$. In the weak-coupling limit $\trel/\tcor\gg 1$, and consequently, the value of the non-Markovianity measure is much greater if the full master equation is used. The effects of the RWA and SA and the calculation of $\N(\Phi)$ will be discussed in more detail in the next section.

\section{Non-Markovian dynamics and the full master equation}
\label{sec_Full_ME}

\subsection{Analytical approximation}
\begin{figure}[!t]
\begin{center}
\includegraphics[scale=0.85]{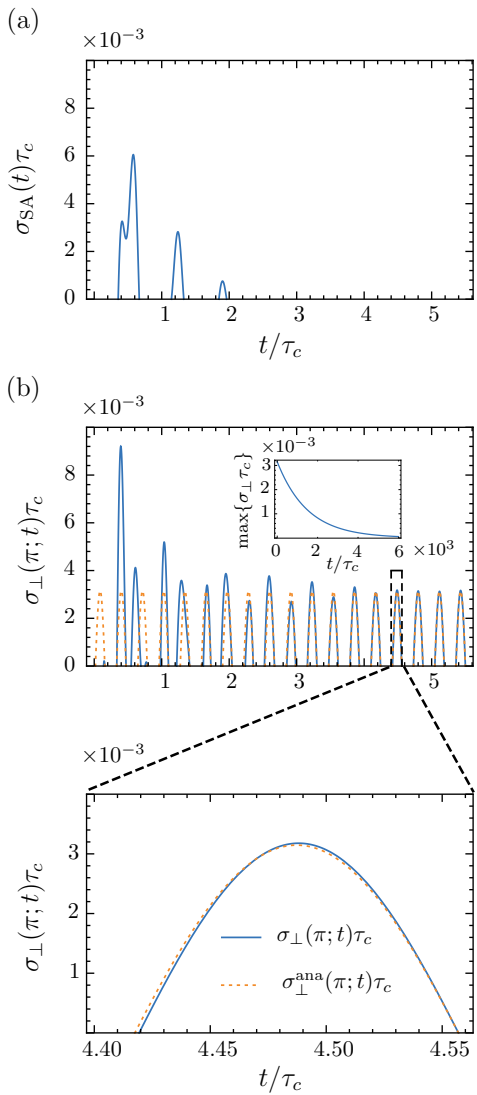}
\end{center}
\caption{(Color online) (a) The positive values of $\sigma$ obtained using the SA master equation and the Lorentzian spectral density. Here $\alpha/\omegaa =0.01$, $\lambda/\omegaa=0.10$, and $\Delta/\omegaa=-0.90$. The initial-state pair corresponds to the one maximizing the non-Markovianity.
(b) As in (a), but for the full master equation. The initial-state pair is that of Eq. \eqref{eq_lambdaIni} with $\xi(0)=\pi$. The solid blue line gives the exact numerical solution $\sigma_\perp$, and the dashed orange line corresponds to the analytical approximation $\sigma_\perp^\text{ana}$ given in Eq.~\eqref{eq_sigmaPerp}.
The bottom panel shows a magnification of part of the top panel. 
The inset gives the maximum value of $\sigma_\perp$ (calculated over one oscillation cycle) as a function of time. It decays on a time scale given by the relaxation time $\trel$. For the chosen parameters $\trel/\tcor\approx 1494$.
\label{fig_sigmaPlot}}
\end{figure}

We start by showing that if the full master equation is used, a good approximation for the value of the non-Markovianity measure can be obtained analytically. The reason for the large reduction in the non-Markovianity following from the use of the SA and RWA (see Fig.~\ref{fig_comparison}) will be explained in the course of the derivation of the analytical formula.
The key observation used in the analytical calculation of the non-Markovianity measure is that if the relaxation time $\trel$ is much longer than the correlation time $\tcor$,
the main contribution to the non-Markovianity measure comes from times $t\gg \tcor$.
In this time domain $\fpm(t)\approx\fpm(\infty),g(t)\approx g(\infty)$, and explicit time dependence only appears in the master equation~\eqref{eq_ME} through the terms $e^{\pm i2\omegaa t}$. These terms lead to a $\sigma(\rhoSI_{1,2};t)$ that oscillates
with frequency $2\omegaa$. In Appendix~\ref{appendix_N} we show that if $\trel\gg\tcor$, the Bloch vectors of a pair of states maximizing the non-Markovianity read
\begin{align}
\label{eq_lambdaIni}
\bflam^1(0)=-\bflam^2(0)=\left(\cos\frac{\xi(0)}{2},\sin\frac{\xi(0)}{2},0\right),
\end{align}
where the optimal angle $\xi(0)\in [0,2\pi)$ has to be determined numerically.
We denote the $\sigma(\hat{\rho}_{1,2};t)$ corresponding to the initial state pair given in  Eq.~\eqref{eq_lambdaIni} by $\sigma_\perp(\xi(0);t)$. In Appendix \ref{appendix_N} we show that $\sigma_\perp(\xi(0);t)\approx\sigma_\perp^{\text{ana}}(\xi(0);t)$, where
\begin{align}
\nonumber
&\sigma_\perp^{\text{ana}}(\xi(0);t)=
\frac{e^{-t/\trel}}{\trel}\\
&\times\left\{\frac{|\h(\infty)|}{\hr(\infty)}\cos[2\omegaa t+\xi(t)+\theta(\infty)]-1\right\}.
\label{eq_sigmaPerp}
\end{align}
Here $\theta(\infty)$ is defined through the equation $g(\infty)=|g(\infty)|e^{i\theta(\infty)}$ and $\xi(t)=\xi(0)-2\hi(\infty)t$. In Fig. \ref{fig_sigmaPlot}(b), we compare $\sigma_\perp^\text{ana}(\xi;t)$ to the exact, numerically calculated value $\sigma_\perp(\xi;t)$.
The agreement between the numerical and analytical results is very good at times $t\gg \tcor$.
In Fig.~\ref{fig_sigmaPlot}(a), we show the time evolution of $\sigma$ obtained using the SA master equation. The oscillations in $\sigma_\text{SA}$ decay on a time scale determined by the correlation time $\tcor$, which is in stark contrast to the long-lived oscillations shown in Fig.~\ref{fig_sigmaPlot}(b). In the latter case the oscillations decay on a time scale given by the relaxation time $\trel$ [see the inset of Fig.~\ref{fig_sigmaPlot}(b)]. For the parameters used in Fig.~\ref{fig_sigmaPlot} we have  $\trel/\tcor \approx 1494$. Consequently, the value of the non-Markovianity measure is much larger if the full master equation is used (see Fig.~\ref{fig_comparison}).

Using $\sigma_\perp^{\text{ana}}$, the time integral appearing
in Eq. \eqref{eq_N} can be calculated analytically. In Appendix \ref{appendix_N} we show that $\N(\Phi)\approx \Nana(\Phi)$, where
\begin{align}
\label{eq_Nana}
\Nana (\Phi) &=  \frac{\nu-\textrm{arctan}(\nu)}{\pi}
\end{align}
and
\begin{align}
\label{eq_nu}
           \nu  &= \frac{|\hi(\infty)|}{\hr(\infty)}.
\end{align}
Equation~\eqref{eq_Nana} has been obtained under the assumptions that $\hr(\infty)>0$ and
$\trel\gg\tcor,\tsys$. Equation~\eqref{eq_Nana} shows that $\Nana (\Phi)$ increases monotonously as $\nu$ grows and that the minimum of $\Nana (\Phi)$, corresponding to Markovian dynamics, is at $\nu=0$.
We will discuss the region of low non-Markovianity occurring near $\nu=0$ in more detail in Sec.~\ref{subsection_low_nonMarkovianity}.

\subsection{Accuracy of the analytical result}
\begin{figure}[t]
\begin{center}
\includegraphics[scale=1.05]{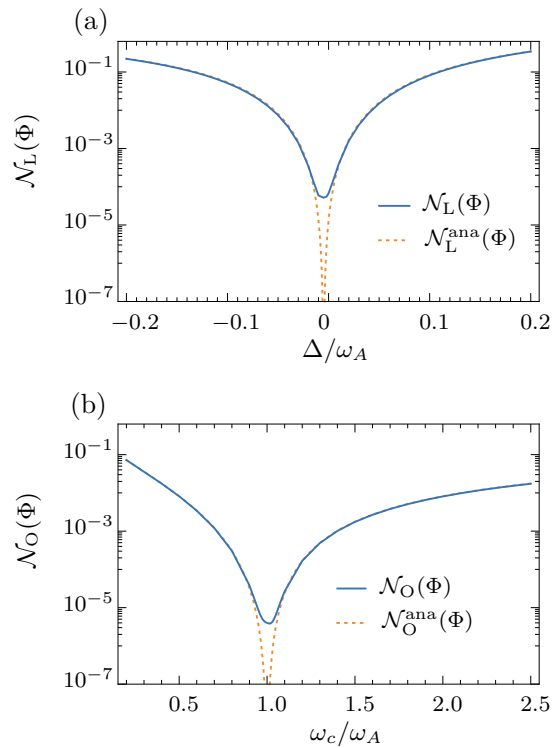}
\end{center}
\caption{(Color online) Non-Markovianity calculated using (a) Lorentzian and (b) Ohmic spectral densities.
The orange dotted line shows the analytical result $\Nana(\Phi)$.
The solid blue  line shows the numerically obtained value of $\N(\Phi)$ corresponding to
 $\alpha/\omegaa=0.01$. In (a), we have chosen $\lambda/\omegaa=\tsys/\trel=0.1$.
\label{fig_dipPlot}}
\end{figure}

In Fig. \ref{fig_dipPlot}, we compare $\Nana(\Phi)$ to the exact,
numerically obtained value $\N(\Phi)$.
The relative error between these two results is very small, except for a narrow region near
the minimum of $\N(\Phi)$.
In this region the non-Markovian dynamics only takes place in the time interval
$[0,t_1]$, where $t_1$ is on the order of the correlation time $\tcor$.
In the derivation of $\Nana(\Phi)$ the non-Markovianity arising in this time interval
has not been taken into account properly, resulting in the observed discrepancy between the numerical and analytical results. Apart from the region of very low non-Markovianity, the analytical result quantifies the value of the non-Markovianity measure extremely accurately.

We have obtained Eq. \eqref{eq_Nana} by taking the limit of vanishing
system-environment interaction $\alpha$, corresponding to $\trel=\infty$ (see Appendix~\ref{appendix_N}).
As shown in  Fig.~\ref{fig_dipPlot}, in the weak-coupling limit the $\alpha$-independent expression $\Nana(\Phi)$ works well. The insensitivity of $\Nana(\Phi)$ to the value of $\alpha$ follows
from the infinite upper bound of the time integration in Eq. \eqref{eq_N} and the fact that $\sigma_\perp$ is proportional to $(1/\trel)e^{-t/\trel}$. When $\trel$ grows
($\alpha$ decreases), the initial amplitude of the oscillations of $\sigma_\perp$ becomes smaller.
This decrease is compensated for by the slower decay of this amplitude.
Due to the infinite upper bound of the time integral,
the net effect of these two changes is that $\Nana(\Phi)$ is almost independent of
$\alpha$ in the weak-coupling limit. This independence is lost if the upper bound of the integration
is finite. In Appendix \ref{appendix_N} we show that the non-Markovianity
arising in the time interval $[0,T]$ is approximately given by
\begin{align}
\label{eq_NanaFiniteT}
\Nana(\Phi;T) = (1-e^{-T/\trel})\ \Nana(\Phi).
\end{align}
This equation has been derived under the assumption that $T\gg\tcor$.
Equation~\eqref{eq_NanaFiniteT} shows that if we decrease the system-environment interaction, and hence increase $\trel$, the amount of non-Markovianity accumulated in the time interval $[0,T]$ becomes smaller. This is what one intuitively expects.

As we argue in Appendix~\ref{appendix_N}, the pair of initial Bloch vectors maximizing the non-Markovianity is of the form given in Eq. \eqref{eq_lambdaIni} regardless of the values of the parameters of the spectral densities. Furthermore, the larger $\nu$ is, the less
the non-Markovianity depends on the angle $\xi(0)$ appearing in $\bflam_\perp^{1,2}(0)$ [see Eq.~\eqref{eq_lambdaIni}]. The $\xi(0)$ dependence is only strong near the minimum of $\N(\Phi)$,  corresponding to $\nu=0$.
The $\xi(0)$ dependence of $\N(\Phi)$ can be explained with the help of Eq.~\eqref{eq_rho10t}. Assume that $\hat{\rho}^j(0)$ is the state corresponding to $\bflam_\perp^j (0)$, $j=1,2$. Thus, 
$\rho_{10}^1(0)=-\rho_{10}^2(0)= 1/2\ e^{-i\xi(0)/2}$, and a straightforward calculation gives
$D[\hat{\rho}^1(t),\hat{\rho}^2(t)]=2|\rho_{10}^1(t)|$. Using Eq.~\eqref{eq_rho10t}, one observes that $|\rho_{10}^1(t)|$, and consequently also $\N(\Phi)$, depends on $\xi(0)$. Note that $v_1(t)=1$ and $v_2(t)=0$ for the dynamical map corresponding to the RWA or SA master equations, indicating that
in these cases $|\rho_{10}^1(t)|$ is independent of $\xi(0)$.

\subsection{Region of low non-Markovianity}
\label{subsection_low_nonMarkovianity}
Figures \ref{fig_comparison} and \ref{fig_dipPlot} show that for both  Lorentzian and Ohmic spectral densities the non-Markovianity is very small in a certain region of the parameter space.  According to Eq.~\eqref{eq_Nana}, this region consists of those parameter values for which  $|\hi(\infty)|/\hr(\infty)\ll 1$.
As we show next, the appearance of a region of low non-Markovianity is a general property of the model considered here.

\begin{figure}[t]
\begin{center}
\includegraphics[scale=1.0]{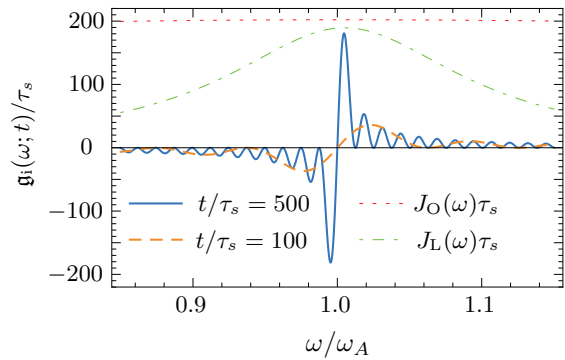}
\end{center}
\caption{
(Color online) The behavior of $\mathfrak{g}_{\text{i}}$ [see Eq.~\eqref{giTimeIntegrated}]
as a function of $\omega$ for two values of $t$ (solid and dashed lines).
For comparison, we also show the behavior of the Ohmic (dotted line) and Lorentzian (dash-dotted line) spectral densities. We have chosen the parameters of $\JO$ and $\JL$ such that the maximum of these functions is at $\omega=\omegaa$. For $\JL$, we have set $\tsys/\tcor=\lambda/\omegaa =0.1$. In the case of $\JL$ and $\JO$, the scale on the vertical axis is arbitrary.
\label{fig_gi}}
\end{figure}

By calculating the time integral appearing in the definition of $\hi(t)$ [see Eq.~\eqref{eq_h}], we obtain $\hi(t) =2\int_0^\infty d\omega J(\omega)\mathfrak{g}_{\text{i}}(\omega;t)$, where
\begin{align}
\label{giTimeIntegrated}
\mathfrak{g}_{\text{i}}(\omega;t) =\frac{\omegaa-\omegaa\cos(\omega t)\cos(\omegaa t)
-\omega\sin(\omega t)\sin(\omegaa t)}{\omega^2-\omegaa^2}.
\end{align}
We plot $\mathfrak{g}_{\text{i}}$ as a function of $\omega$ for two values of $t$ in Fig.~\eqref{fig_gi}.
For $t\gg\tsys$ and $\omega$ such that $|\omega-\omegaa|/\omegaa \ll 1$, $\mathfrak{g}_{\text{i}}$ is approximately an odd function of $\omega-\omegaa$. Furthermore, the oscillation amplitude of $\mathfrak{g}_{\text{i}}$ decays fast with increasing $|\omega-\omegaa|$.
Using these properties of $\mathfrak{g}_{\text{i}}$ and assuming that the spectral density $J$ is an even function of $\omega-\omegaa$ near $\omega=\omegaa$, we find that $\hi(t\gg\tcor)\approx 0$.
$J$ can typically be assumed to be an even function of $\omega-\omegaa$ near $\omega=\omegaa$ if the  maximum of $J$ is at $\omega=\omegaa$. Hence the dynamics is nearly Markovian if the spectral density is maximized at a frequency that corresponds to the energy
separation of the two-level system. A similar result was obtained earlier in Ref.~\cite{Clos12}.

We show next that for the spectral densities used here, this condition for the appearance of Markovian dynamics works well. The maximum of $\JO$
is at $\omega=\omegaa$ if $\omegac=\omegaa$. Setting $\omegac$ equal to $\omegaa$ in the expression for $\hi(\infty)$ corresponding to the Ohmic spectral density [see~\eqref{eq_hOh}] gives $\hi(\infty)=0$,
which is in perfect agreement with the above argumentation. The maximum of the Lorentzian spectral density is at $\omega=\omegaa$ if $\Delta=0$. For this value of $\Delta$, the equations given in Appendix~\ref{appendix_Functions} yield $|\hi(\infty)|/\hr(\infty)<\lambda/2\omegaa =\tsys/2\tcor$. In this paper, we assume that  $\tsys/\tcor\ll 1$ [see the discussion following Eq.~\eqref{eq_JO}], so that  the predicted result is obtained also in this case.

\subsection{Secular and Markov approximations}
It is typically assumed that the SA is justified if the time scale of the intrinsic dynamics $\tsys$
is much shorter than any other time scale in the problem. In the system studied here, however,  $\N(\Phi)$ and $\N(\PhiSA)$ do not, in general, approach each other even if $\tsys\to 0$.
With the help of Eqs. \eqref{eq_Nana}, \eqref{eq_nu}, and \eqref{eq_hOh} we find that $\Nana_\text{\text{O}}(\Phi)$
increases with decreasing $\tsys$.
On the other hand, $\N_\text{O}(\PhiSA)=\N_\text{O}(\PhiRWA)=0$ for any value of $\tsys$. 
We stress that although the RWA and SA may lead to accurate description of the density operator, non-Markovianity is described by a temporal integral, and hence apparently negligible errors in the state may accumulate and have serious consequences. This type of behavior is similar to the general framework considered in Ref.~\cite{Salmilehto12}, in which the SA has been observed to lead to contradicting results for a temporal integral of operator current and the corresponding temporal change in the expectation value of the operator.

The Markov approximation is commonly used in the derivation of Lindblad-form master equations~\cite{Breuer}. In our case, it corresponds to extending the upper bound of the integral over $s$ to infinity in Eqs. \eqref{eq_fpm} and \eqref{eq_h}. In other words,  $\fpm(t)$  is replaced with $\fpm(\infty)$ and $\h(t)$ with $\h(\infty)$.  The same replacements are used in the derivation of $\Nana(\Phi)$. It follows that the Markov approximation leaves the non-Markovianity unaffected as long as $\Nana(\Phi)$ can be assumed to be a good estimate for $\N(\Phi)$.  In the case of the RWA and SA master equations the situation is different; the dynamics becomes Markovian if the Markov approximation is used.

\section{Conclusions}
\label{sec_conclusions}
The main result of this paper is that the RWA and SA can lead to a dramatic reduction in the amount of non-Markovianity detected by a commonly used non-Markovianity measure. Hence, although the RWA and SA are often considered to be justifiable approximations, they should be used with caution if quantifying non-Markovianity  is of interest. We obtained this result by considering an open quantum system consisting of a two-level system interacting weakly with a bosonic zero-temperature environment. We showed that the rapidly oscillating terms that are neglected in the RWA and SA contribute to non-Markovian dynamics during the whole relaxation time of the system, whereas if the RWA or SA is used, non-Markovian dynamics takes place only in a time interval whose length is on the order of the reservoir correlation time $\tcor$.
In the limit of a weak system-environment interaction,
the correlation time is very short in comparison with the relaxation time. Consequently,
the amount of non-Markovianity can be orders of magnitude greater if the full master equation is used instead of the RWA or SA master equation.

Without employing the RWA and SA, we derived an
analytical expression for the non-Markovianity measure for a general spectral density. Remarkably, this expression can be given
in terms of the asymptotic $t\to\infty$ values of the coefficients of the master equation, without
solving the dynamics of the system.
Since knowing the asymptotic values is sufficient, the Markov approximation has a negligible effect on the non-Markovianity if neither the RWA nor the SA is employed.
We confirmed the validity of the analytical formula for non-Markovianity numerically.
With the help of this analytical result,
we showed that, generally, the Lorentzian and Ohmic spectral densities lead to nonvanishing non-Markovianity. However, the dynamics is approximately Markovian if the value of the spectral density is maximized at a frequency corresponding to the
energy separation of the two-level system. We argued that this a general phenomenon
that occurs for a large class of spectral densities. We also found that the value of the non-Markovianity measure depends only very weakly on the strength of the system-environment interaction in the weak-coupling limit. We showed that this results from the infinite upper bound
of the time integral appearing in the definition of the non-Markovianity measure.

We argued that the pair of initial states maximizing the non-Markovianity
corresponds to two antipodal points on the equator of the Bloch sphere.
This is in contrast to the case of the RWA and SA master equations that
 lead to maximal non-Markovianity if the points are at the poles of the Bloch sphere.
Furthermore, the non-Markovianity of the dynamics depends on the initial phases of the off-diagonal
elements of the density operators.

In the future, it will be interesting to apply the analytical approach
presented in this paper to other models where the RWA and SA have not been used.
For example, it should be possible to reproduce the numerical results of Ref. \cite{Clos12}
using analytical means.

\acknowledgments
We thank A. Karlsson, E.-M. Laine, K. Luoma, and S. Maniscalco for illuminating discussions and J. Piilo for comments on the manuscript.
This research has been supported by the Alfred \mbox{Kordelin} Foundation and the Academy of Finland through its Centres of Excellence Program (Project No. 251748) and Grants No.~138903 and No.~135794. 

\appendix

\section{COEFFICIENTS OF THE MASTER EQUATION}
\label{appendix_Functions}
For the Lorentzian spectral density, the functions appearing in the master equation \eqref{eq_ME} read
\begin{align}
\nonumber
\label{eq_fmLo}
\fm(t) =& 2\pi \JL(\omegaa) \\
&\times\left\{1-e^{-\lambda t}
\left[\cos(\Delta t)-\frac{\Delta}{\lambda} \sin(\Delta t)\right]\right\},\\
\label{eq_fpLo}
\fp(t) =& \fm(t)\Big|_{\Delta\to\Delta-2\omegaa},\\
\label{eq_hiLo}
\hi(\infty) = &\pi
\left[-\frac{\Delta}{\lambda}\JL(\omegaa)+\frac{\Delta-2\omegaa}{\lambda}\JL(-\omegaa)\right],
\end{align}
and $\hr(t)=[\fm(t)+\fp(t)]/2$. The expression for $\hi(t)$ is very long and hence is not shown here.
In the calculation of Eqs. \eqref{eq_fmLo}--\eqref{eq_hiLo}, we have extended the lower bound of the frequency integral to $-\infty$. This can be done if $\JL(0)$ is much smaller than the maximum value $\alpha/2\pi$ of $\JL$, so that the contribution to the integrals from the frequency range $(-\infty,0)$
is small compared with the contribution arising from the range $[0,\infty)$. It is straightforward to see that $\JL(0)\ll \alpha/2\pi$ if $\omegaa-\Delta\gg\lambda.$ 
The expressions for $\fpm(t)$ and $\h(t)$ corresponding to the Ohmic spectral density are very long
and are therefore not given here. The asymptotic values of these functions read
\begin{align}
&\fm(\infty) =  2\pi \JO (\omegaa),\\
&\fp(\infty)\ll \fm(\infty),\\
\label{eq_hOh}
&\h(\infty) =\pi \JO(\omegaa)\left[1-i\frac{2}{\pi}\ln\left(\frac{\omegaa}{\omegac}\right)
\right].
\end{align}

\section{DYNAMICAL MAP}
\label{appendix_Dynamical_map}
The operators $\{\hat{A}_j\}$ appearing in Eq.~\eqref{eq_dynmap} are defined as
\begin{align}
\hat{A}_j(t) =
\begin{cases}
\displaystyle\frac{w_j(t)\mathbb{I}_2+\sigz}{\sqrt{1+|w_j(t)|^2}},\quad &j=1,2,\\
\displaystyle\frac{i w_j(t)\sigx+\sigy}{\sqrt{1+|w_j(t)|^2}}, &j=3,4,
\end{cases}
\end{align}
where 
\begin{align}
w_j(t) =
\begin{cases}
\displaystyle\frac{(-1)^j\sqrt{\frac{1}{4}u^2(t)+|p_1(t)|^2}+p_{1,\text{r}}(t)}{\frac{1}{2}u(t)
+i p_{1,\text{i}}(t)}, &j=1,2,\\
\displaystyle\frac{(-1)^j\sqrt{\frac{1}{4}u^2(t)+|p_2(t)|^2}+p_{2,\text{r}}(t)}{\frac{1}{2}u(t)+i p_{2,\text{i}}(t)},&j=3,4,
\end{cases}
\end{align}
and $p_k(t) =p_{k,\text{r}}(t)+ip_{k,\text{i}}(t)= e^{-\Gamma(t)}v_k^*(t),\quad k=1,2.$

\section{CALCULATION OF THE NON-MARKOVIANITY MEASURE}
\label{appendix_N}
\subsection{RWA and SA master equations}
We consider first the SA master equation.
We denote the Bloch vector of $\rhoSI_j$ by $\boldsymbol{\lambda}^j$ and define
\begin{align}
\label{eq_deltalambda}
\delta\bflam(t)= \bflam^1(t)-\bflam^2(t).
\end{align}
Using this, we obtain
\begin{align}
\label{eq_sigmalambda}
\sigma(\rhoSI_{1,2};t) &=\frac{d}{dt} \frac{1}{2} \|\delta\bflam(t)\|,
\end{align}
where $\|\delta\bflam(t)\|=\sqrt{\delta\bflam(t)\cdot \delta\bflam(t)}$. The solution of the SA master equation can be obtained by setting $\Gammai(t)=0$, $v_1(t)=1$, and $v_2(t)=0$ in Eqs. \eqref{eq_rho10t} and \eqref{eq_rho01t}.
This gives $\delta\lambda_{x,y}(t)=e^{-\Gammar(t)/2}\delta\lambda_{x,y}(0)$ and
$\delta\lambda_{z}(t)=e^{-\Gammar(t)}\delta\lambda_{z}(0)$. Using these equations and Eq.~\eqref{eq_sigmalambda}, we obtain
\begin{align}
&\sigma_{\text{SA}}(\delta\bflam(0);t)= -\frac{\hr(t)}{2}
\|\delta\bflam(t)\|
\left[ 1+\frac{e^{-2\Gammar(t)}\delta\lambda_z(0)^2}{\|\delta\bflam(t)\|^2}\right],
\label{sigmaSec}
\end{align}
where
\begin{align}
\|\delta\bflam(t)\|^2=e^{-\Gammar(t)}
[\|\delta\bflam (0)\|^2 +(e^{-\Gammar(t)}-1)\delta\lambda_z(0)^2 ].
\end{align}
Only positive values of $\sigma_{\text{SA}}$ contribute to $\N(\PhiSA)$. As is evident from
Eq.~\eqref{sigmaSec}, $\sigma_{\text{SA}}(\delta\bflam(0);t)$ is only positive if $\hr(t) <0$. In Ref.~\cite{Wissmann12}, it has been shown  that the initial-state pair
maximizing the non-Markovianity corresponds to two antipodal points on the Bloch sphere.
This implies that $\|\delta\bflam(0)\|=2$.
For any fixed $t$ yielding a negative $\hr(t)$, the value of $\sigma_{\text{SA}}(\delta\bflam(0);t)$ is maximized either at $|\delta\lambda_z(0)|=0$, at $|\delta\lambda_z(0)|=2$, or at $|\delta\lambda_z(0)|=\delta\lambda_{z;0}$,  where $\delta\lambda_{z;0}$ is a point where the derivative of  $\sigma_{\text{SA}}(\delta\bflam(0);t)$  with respect to  $|\delta\lambda_z (0)|$ vanishes.
Straightforward calculation shows that $|\delta\lambda_z (0)|=2$ is the correct choice irrespective of the value of $t$. The non-Markovianity is thus maximized by
\begin{align}
&\sigma_{\text{SA}}(|\delta\lambda_z(0)|=2;t)= -2\hr(t)e^{-\Gammar(t)}.
\end{align}
In the case of the RWA master equation, the calculation proceeds similarly,
with the exception that $2\hr=\fp+\fm$ is replaced with $\fm$ and $\Gammar$ with
$\Gamma^{\textrm{RWA}}$.

\subsection{Full master equation}
In terms of the components of the Bloch vector, the master equation~\eqref{eq_ME} reads
\begin{align}
\label{dotlx}
\nonumber
\dot{\lambda}_{x}(t) =& -\hr(t)\lambda_{x}(t)
+ |\h(t)|\cos[2\omegaa t+\theta(t)]\lambda_{x}(t)\\
& -2|\h(t)|\cos[\omegaa t+\theta(t)]\sin(\omegaa t)\lambda_{y}(t),\\
\label{dotly}
\nonumber
\dot{\lambda}_{y}(t) =& -\hr(t)\lambda_{y}(t)
-|\h(t)|\cos[2\omegaa t+\theta(t)]\lambda_{y}(t)\\
& -2|\h(t)|\sin[\omegaa t+\theta(t)]\cos(\omegaa t)\lambda_{x}(t),\\
\label{dotlz}
\dot{\lambda}_z(t) =&\fp(t)-\fm(t) -2\hr(t) \lambda_z(t),
\end{align}
where the angle $\theta$ is defined through the equation $\h(t)=|\h(t)|e^{\theta(t)}$.
With the help of these equations and Eqs. \eqref{eq_deltalambda} and \eqref{eq_sigmalambda}, we have
\begin{align}
\nonumber
\label{eq_sigmaExact}
&\sigma(\rhoSI_{1,2};t)=
\frac{\|\delta\bflam(t)\|}{2}\bigg\{-\hr(t)\left[1+\frac{\delta\lambda_z(t)^2}
{\|\delta\bflam(t)\|^2}\right]\\
&+|g(t)|\cos[2\omegaa t+\theta(t)+\xi(t)] \left[1-\frac{\delta\lambda_z(t)^2}
{\|\delta\bflam(t)\|^2}\right]\bigg\},
\end{align}
where
\begin{align}
\xi(t) &=\text{sgn}[\delta\lambda_x(t)\delta\lambda_y(t)]\arccos\left[\frac{\delta\lambda_x(t)^2-\delta\lambda_y(t)^2}{\delta\lambda_x(t)^2+\delta\lambda_y(t)^2}\right].
\end{align}
We discuss next the initial-state pair maximizing the non-Markovianity.
We assume that $\hr(\infty)>0$ and divide the discussion into two parts, based on the value of the quantity
\begin{align}
\label{eq_mu}
\x=\frac{|\h(\infty)|}{\hr(\infty)}\geq 1.
\end{align}
\subsubsection{The case of $\x =1$}
Assume first that $\x = 1$. In this case Eq. \eqref{eq_sigmaExact} is negative or zero
if $t\gg\tcor$. Non-Markovian dynamics is thus possible only during a time interval
$[0,t_1]$, where $t_1$ is on the order of the correlation time. For the Ohmic
spectral density $\hr(t)$ is always non-negative, and consequently, the first
term inside the curly braces in Eq. \eqref{eq_sigmaExact} is maximized by choosing $\delta\lambda_z(0)=0$, so that $\delta\lambda_z(t)=0$.
The same choice maximizes the second term inside the braces [we assume that $t$ is such that this term is positive; if it were negative, $\sigma(t)$ would also be negative].
Using Eqs. \eqref{eq_rho11t}--\eqref{eq_rho01t}, we find that $\delta\lambda_{x,y}(t)\propto e^{-\Gammar(t)/2}$ and $\delta\lambda_{z}(t)\propto e^{-\Gammar(t)}$.
Because $\delta\lambda_z$ decays faster than $\delta\lambda_{x,y}$, the overall
factor in Eq. \eqref{eq_sigmaExact} is also maximized by setting $\delta\lambda_z(0)=0$.
In conclusion, as the largest possible value for the norm is $\|\delta\bflam(0)\|=2$, the non-Markovianity is maximized by choosing $\delta\bflam(0)=\delta\bflam_\perp(0)$,
where
\begin{align}
\delta\bflam_\perp(0)=2\left(\cos\frac{\xi(0)}{2},\sin\frac{\xi(0)}{2},0\right),\quad \xi(0)\in [0,2\pi).
\label{eq_xi}
\end{align}
For the Lorentzian spectral density $\hr(t)$ can be positive or negative,
depending on the values of the parameters of the spectral density. However, we checked numerically that $\hr(t)$ is non-negative  for those parameter values for which $\x=1$. As a consequence, an 
initial-state pair of the form given in Eq. \eqref{eq_xi} maximizes the non-Markovianity also in this case.

We denote by $\sigma_\perp(\xi(0);t)$ the value of $\sigma(\rhoSI_{1,2};t)$ obtained by choosing the initial states as in Eq. \eqref{eq_xi},
\begin{align}
\label{eq_sigmaPerpExact}
\nonumber
&\sigma_\perp(\xi(0);t)=-\frac{\|\delta\bflam_\perp(t)\|}{2}\\
&\times\bigg\{\hr(t)-|\h(t)|\cos[2\omegaa t+\theta(t)+\xi(t)]\bigg\}.
\end{align}
For both spectral densities, the value of the angle $\xi(0)$ maximizing the non-Markovianity has to be determined numerically.

\subsubsection{The case of $\x > 1$}
If $\x > 1$, non-Markovian dynamics is also possible if $t\gg\tcor$.
It is straightforward to show that the non-Markovianity is maximized  in the time domain $t\gg\tcor$  by the initial-state pair given in Eq. \eqref{eq_xi}. For a general spectral density, this state pair does not necessarily maximize the non-Markovianity
in the time domain $[0,t_1]$, where $t_1$ is on the order of $\tcor$. However,
if the relaxation time $\trel$ is much longer than the correlation time $\tcor$,
the contribution to the integral in Eq. \eqref{eq_N} from the time interval $[0,t_1]$ can be assumed to be negligible compared to the contribution resulting from times $t\gg \tcor$. The initial-state pair maximizing the non-Markovianity can thus be assumed to be of the form shown in Eq. \eqref{eq_xi}.

For $t$ much greater than the correlation time, we have $\fpm (t)\approx\fpm (\infty)$,
$|\h(t)|\approx |\h(\infty)|$, $\theta(t)\approx\theta(\infty)$, and $\Gamma(t)\approx 2\h(\infty)t$.  Furthermore, $v_1(t)\approx 1$ and $v_2(t)\approx 0$. These approximations yield
\begin{align}
\label{eq_perpt}
\delta\bflam_\perp(t)\approx 2e^{-t/\trel}\left(\cos\frac{\xi(t)}{2},\sin\frac{\xi(t)}{2},0\right),
\end{align}
where $\xi(t)= \xi(0)-2\hi (\infty)t$. We denote the  $\sigma$ obtained using Eq. \eqref{eq_perpt} by
$\sigma_\perp^\text{ana}$. Equations~\eqref{eq_sigmaPerpExact} and~\eqref{eq_perpt} give
\begin{align}
\label{eq_sigmaanaapp}
& \sigma_\perp^\text{ana}(\xi(0);t)= \frac{e^{-t/\trel}}{\trel}
\left\{\x\cos\left[2\omegaa t+\theta(\infty) + \xi(t)\right]-1\right\}.
\end{align}
The value of Eq. \eqref{eq_sigmaanaapp} is positive for every $t\in (t_n^{-},t_n^{+})$, where
\begin{align}
\label{eq_tnpm}
t_n^{\pm} &=\frac{2n\pi-\theta(\infty)-\xi(0)\pm \arccos\left(\frac{1}{\x}\right)}
{2[\omegaa-\hi (\infty)]}.
\end{align}
and $n=0,1,2,\ldots$. Using Eqs. \eqref{eq_sigmaanaapp} and \eqref{eq_tnpm}, we find that
\begin{align}
\nonumber
&\frac{1}{2}\int_{0}^{\infty} dt\ [|\sigma_\perp^\text{ana} (t)|+\sigma_\perp^\text{ana} (t)]
=e^{\epsilon [\pi+\theta(\infty)+\xi(0)]}\\
&\times \frac{\epsilon\cosh[\epsilon\ \text{arcsec}(\x)]\sqrt{\x^2-1}-\sinh[\epsilon\ \text{arcsec}(\x)]}{\sinh(\epsilon\pi)(1+\epsilon^2)}\\
&= \frac{\sqrt{\x^2-1}-\textrm{arcsec}(\x)}{\pi}\left\{1+[\pi+\theta+\xi(0)]\epsilon\right\}+\mathcal{O}(\epsilon^2),
\label{eq_NanaExact}
\end{align}
where
\begin{align}
\epsilon =\frac{1}{2\trel[\omegaa-\hi (\infty)]}\propto \frac{\alpha}{\omegaa}.
\end{align}
In the case of a very weak system-environment interaction
(or, equivalently, $\trel\gg\tsys$), it is enough to include only the lowest-order term
in Eq. \eqref{eq_NanaExact}.
This gives  $\N(\Phi)\approx \Nana(\Phi)$, where
\begin{align}
\label{eq_Nana_appen}
\Nana(\Phi) =
\displaystyle\frac{\sqrt{\x^2-1}-\textrm{arcsec}(\x)}{\pi}.
\end{align}
Note that this equation is independent of the angle $\xi(0)$ appearing in Eq. \eqref{eq_xi}.
Defining $\nu=|\hi(\infty)|/\hr(\infty)$, we can write $\Nana(\Phi)$ in the form given by   Eq.~\eqref{eq_Nana}.

It may sometimes be of interest to quantify the non-Markovianity accumulated during a time interval $[0,T]$, where $T$ is finite. On the lines of the above calculation, we obtain
\begin{align}
\Nana(\Phi;T) = (1-e^{-T/\trel})\ \Nana(\Phi),
\end{align}
where $T\gg\tcor$.
In obtaining this equation we have taken the limit $\epsilon\to 0$,
keeping $T/\trel$ fixed.


\begin{thebibliography}{33}

\bibitem{Breuer} H.-P. Breuer and F.~Petruccione, \emph{The Theory of Open Quantum Systems}
(Oxford University Press, Oxford, 2007).

\bibitem{Weiss} U.~Weiss, \emph{Quantum Dissipative Systems}, (World Scientific, Singapore, 2008).

\bibitem{Wolf08} M. M. Wolf, J. Eisert, T. S. Cubitt, and J. I. Cirac, Phys. Rev. Lett. {\bf 101}, 150402 (2008).

\bibitem{Breuer09} H.-P. Breuer, E.-M. Laine, and J. Piilo, Phys. Rev. Lett. {\bf 103}, 210401 (2009).

\bibitem{Rivas10} \'A. Rivas, S. F. Huelga, and M. B. Plenio, Phys. Rev. Lett. {\bf 105}, 050403 (2010).

\bibitem{Laine10} E.-M. Laine, J. Piilo, and H.-P. Breuer, Phys. Rev. A {\bf 81}, 062115 (2010).

\bibitem{Chruscinski10}  D. Chru\'{s}ci\'{n}ski and A. Kossakowski, Phys. Rev. Lett. {\bf 104}, 070406 (2010).

\bibitem{Lu10} X.-M. Lu, X. Wang, and C. P. Sun, Phys. Rev. A {\bf 82}, 042103 (2010).

\bibitem{Chruscinski11} D. Chru\'{s}ci\'{n}ski, A. Kossakowski, and A. Rivas, Phys. Rev. A {\bf 83}, 052128 (2011).

\bibitem{Apollaro11} T. J. G. Apollaro, C. Di Franco, F. Plastina, and M. Paternostro, Phys. Rev. A {\bf 83}, 032103 (2011).

\bibitem{Haikka11} P. Haikka, J.D. Cresser, and S. Maniscalco, Phys. Rev. A {\bf  83}, 012112 (2011).

\bibitem{Znidaric11} M. \v{Z}nidari\v{c}, C. Pineda, and I. Garc\'{i}a-Mata, Phys. Rev. Lett. {\bf 107}, 080404 (2011).

\bibitem{Liu11} B.H. Liu, L. Li, Y.-F. Huang, C.-F. Li, G.-C. Guo, E.-M. Laine, H.-P. Breuer, and J. Piilo, Nat. Phys. {\bf 7}, 931 (2011).

\bibitem{Mazzola12} L. Mazzola, C. A. Rodr\'iguez-Rosario, K. Modi, and M. Paternostro, Phys. Rev. A {\bf 86}, 010102(R) (2012).

\bibitem{Rodriguez12}  C. A. Rodr\'iguez-Rosario, K. Modi, L. Mazzola, and A. Aspuru-Guzik, EPL {\bf 99}, 20010 (2012).

\bibitem{Clos12} G. Clos and H.-P. Breuer, Phys. Rev. A {\bf 86}, 012115 (2012).

\bibitem{Zeng12} H. S. Zeng, N. Tang, Y. P. Zheng, and T. T. Xu,
Eur. Phys. J. D {\bf 66}, 255 (2012).

\bibitem{Zhang12} W.-M. Zhang, P.-Y. Lo, H.-N. Xiong, M. W.-Y. Tu, and F. Nori, Phys. Rev. Lett. {\bf 109}, 170402 (2012).

\bibitem{Wissmann12} S. Wi\ss mann, A. Karlsson, E.-M. Laine, J. Piilo, and H.-P. Breuer,
Phys. Rev. A {\bf 86}, 062108 (2012).

\bibitem{Bylicka13} B. Bylicka, D.  Chru\'{s}ci\'{n}ski, and S. Maniscalco,
arXiv:1301.2585.

\bibitem{Walls} D. F. Walls and G. J. Milburn, \emph{Quantum Optics} (Springer, Berlin, 1994).

\bibitem{Scully} M. O. Scully and M. S. Zubairy, \emph{Quantum Optics}
(Cambridge University Press, Cambridge, 1997).

\bibitem{Agarwal} G. S. Agarwal, \emph{Quantum Statistical Theories of Spontaneous Emission and Their Relation to Other Approaches}
(Springer, Berlin, 1974).

\bibitem{Agarwal71} G. S. Agarwal, Phys. Rev. A {\bf 4}, 1778 (1971).

\bibitem{Agarwal73} G. S. Agarwal, Phys. Rev. A {\bf 7}, 1195 (1973).

\bibitem{Ford97} G. W. Ford and R. F. O'Connell, Phys. A {\bf 243}, 377 (1997).

\bibitem{Intravaia03} F. Intravaia, S. Maniscalco, and A. Messina,
Eur. Phys. J. B {\bf 32}, 97 (2003).

\bibitem{Zheng08} H. Zheng, S. Y. Zhu, and M. S. Zubairy, Phys. Rev. Lett. {\bf 101}, 200404 (2008).

\bibitem{Larson12} J. Larson, Phys. Rev. Lett. {\bf 108}, 033601 (2012).

\bibitem{Peano12} V. Peano, M. Marthaler, and M. I. Dykman, Phys. Rev. Lett. {\bf 109}, 090401 (2012).

\bibitem{Pekola10} J. P. Pekola, V. Brosco, M. M\"ott\"onen, P. Solinas, and A. Shnirman,
Phys. Rev. Lett. {\bf 105}, 030401 (2010).

\bibitem{Solinas10} P. Solinas, M. M\"ott\"onen, J. Salmilehto, and J. P. Pekola,
Phys. Rev. B {\bf 82}, 134517 (2010).

\bibitem{Salmilehto12} J. Salmilehto, P. Solinas, and M. M\"ott\"onen,
Phys. Rev. A {\bf 85}, 032110 (2012).

\bibitem{Benatti10} F. Benatti, R. Floreanini, and U. Marzolino, Phys. Rev. A {\bf 81}, 012105 (2010).

\bibitem{Fleming10} C. Fleming, N. I. Cummings, C. Anastopoulos, and B. L. Hu,
J. Phys. A {\bf 43}, 405304 (2010).

\bibitem{footnote}
The value of the non-Markovianity measure used in this paper is invariant in transformations of the form $\hat{\rho}(t)\to \hat{\tilde{\rho}}(t)=\hat{U}(t)\hat{\rho}(t)\hat{U}^\dagger (t)$,
where $\hat{U}(t)$ is unitary. This invariance can be used to eliminate the first term on the right hand side of the master equations~\eqref{eq_ME}, \eqref{eq_ME_RWA}, and~\eqref{eq_ME_SA}.
By defining $\hat{U}(t)=e^{-i\delta(t)\frac{1}{2}\sigz}$, where $\delta(t)=\int_{0}^{t}ds\ h(s)$,
the master equation~\eqref{eq_ME_RWA} becomes
$\frac{d}{dt}\hat{\tilde{\rho}}(t)=\fm(t) [\sigm\hat{\tilde{\rho}}(t)\sigp
-\frac{1}{2}\{\sigp\sigm,\hat{\tilde{\rho}}(t)\}$]. Similarly, by replacing $h$ with $\hi$ in the definition of $\delta$, the SA master equation [Eq.~\eqref{eq_ME_SA}] can be cast in the form
 $\frac{d}{dt}\hat{\tilde{\rho}}(t)=\sum_{k=\pm}f_k(t) [\sig_k\hat{\tilde{\rho}}(t)\sig_k^\dagger
-\frac{1}{2}\{\sig_k^\dagger\sig_k,\hat{\tilde{\rho}}(t)\}]$ and the full master equation [Eq.~\eqref{eq_ME}] becomes $\frac{d}{dt}\hat{\tilde{\rho}}(t)=\sum_{k=\pm}f_k(t) [\sig_k\hat{\tilde{\rho}}(t)\sig_k^\dagger
-\frac{1}{2}\{\sig_k^\dagger\sig_k,\hat{\tilde{\rho}}(t)\}]+e^{2i[\omegaa t-\delta(t)]} \h(t)\sigp\hat{\tilde{\rho}}(t)\sigp+e^{-2i[\omegaa t-\delta(t)]} \h^*(t)\sigm\hat{\tilde{\rho}}(t)\sigm$.
\end{thebibliography}
\end{document}